\def\vec#1{{\rm\bf #1}}

\documentclass[fleqn,12pt,twoside]{article}
\usepackage{espcrc1}

\usepackage{epsfig}

\usepackage[figuresright]{rotating}

\newcommand{\AmS}{{\protect\the\textfont2
  A\kern-.1667em\lower.5ex\hbox{M}\kern-.125emS}}

\title{A basis for the statistical mechanics of granular systems}

\author{
Sam. F. Edwards \address[MCSD]{ Polymers and Colloids Group, Cavendish 
Laboratory,
University of
Cambridge,\\ Madingley Road, Cambridge CB3 0HE, UK}, 
       Jasna Bruji\'c {\addressmark[MCSD]},
       Hern\'an A. Makse \address{ Levich Institute and Physics Department, 
        City College of New York,\\
New York, NY 10031, US}
}       
\begin{document}

\maketitle

\begin{abstract}

This paper aims to justify the use of statistical
mechanics tools in situations where the system is out of
equilibrium and jammed. Specifically, we derive a Boltzmann equation for a jammed granular system and show that the Boltzmann's analysis can be used to produce a
``Second Law'', $\dot{S}\ge 0$ for jammed systems.
We highlight the fundamental questions in
this area of physics and point to the key quantities in
characterising a packing of particles, accessible through a novel
experimentation method which we also present here.
\end{abstract}

\section{Introduction}

In a thermal system, the Brownian motion of the constituent
particles implies that the system dynamically explores the
available energy landscape, such that the notion of a statistical
ensemble applies. For densely packed systems of interest in this
study, in which enduring contacts between particles are important,
the potential energy barrier prohibits an equivalent random
motion. At first sight it seems that the thermal statistical
mechanics do not apply to these systems as there is no mechanism
for averaging over the configurational states. Hence, these
systems are inherently out of equilibrium. On the other hand, if
the granular material is gently tapped such that the grains can
slowly explore the available configurations, the situation becomes
analogous to the equilibrium case scenario. It has been shown that
the volume of the system is dependent on the applied tapping
regime, and that this dependence is reversible, implying
ergodicity \cite{chicago2}. This result gives support to the
proposed statistical ensemble valid for dense, static and slowly
moving granular materials which was first introduced by Edwards
and Oakeshott in 1989 \cite{edwards3,edwards2}. Through this approach, notions
of macroscopic quantities such as entropy and compactivity were
also introduced to granular matter.

Here we present a theoretical framework
to fully describe the exact specificities of  granular packings,
and a shaking scenario which leads to  
the derivation of the 
Boltzmann equation for a jammed granular system.
This kind of an analysis paves the path to the study of 
macroscopic quantities,
such as  the compactivity, characterising each jammed
configuration from the microstructural information of the packing.
It is according to this theory that the static configurations
obtained from experiments are later characterised.
An extended version of this paper is presented in \cite{mbe}.

\section{Classical Statistical Mechanics}

We first present the classical statistical mechanics theorems 
to an
extent which facilitates an understanding of the important
concepts for the development of an analogous granular theory, as
well as the assumptions necessary for the belief in such a
parallel approach.
In the conventional statistical mechanics of thermal systems, the
different possible configurations, or microstates, of the system
are given by points in the phase space
of all positions and momenta
\{$p,q$\} of the constituent particles.
The equilibrium 
probability density
$\rho_{\mbox{\scriptsize eqm}}$ must be a stationary state of  
Liouville's equation
which implies that 
$\rho_{\mbox{\scriptsize eqm}}$
 must be expressed only in
terms of 
the total energy of the system, $E$.
The simplest form for a system with Hamiltonian ${\cal H}(p,q)$
is
the microcanonical distribution:
\begin{equation}
\rho_{\mbox{\scriptsize eqm}}(E) = \frac{1}
{\Sigma_{\mbox{\scriptsize eqm}}(E)},
\label{equal}
\end{equation}
for the microstates within the ensemble, ${\cal H}(p,q)=E$, 
and zero otherwise.
Here,
\begin{equation} 
\Sigma_{\mbox{\scriptsize eqm}}(E) = \int
\delta(E - {\cal H}(p,q) ) ~~ dp ~ dq,
\end{equation}
is the area of energy surface ${\cal H}(p,q)=E$. 

Equation (\ref{equal}) states that all microstates are equally probable.
Assuming that
this is the true distribution of the system implies accepting the
ergodic hypothesis, i.e. the trajectory of the closed system will
pass arbitrarily close to any point in phase space.

It was the remarkable step of Boltzmann to associate this
statistical concept of the number of microstates with the
thermodynamic notion of entropy through his famous formula
\begin{equation}
S_{\mbox{\scriptsize eqm}}(E) = k_B \log \Omega{\mbox{\scriptsize
eqm}}(E).
\end{equation}



Whereas the study of thermal systems has had the advantage of
available statistical mechanics tools for the exploration of the
phase space, an entirely new statistical method, unrelated to the
temperature, had to be constructed for grains.

\section{Statistical Mechanics for Jammed Matter}

We now  consider a jammed granular system composed of {\it rigid}
grains.
Such a system is analogously described by a network of contacts
between the constituent particles in a fixed volume $V$, since
there is no relevant energy $E$ in the system. In the case of
granular materials, the analogue of phase space, the space of
microstates of the system, is the space of possible jammed
configurations as a function of the degrees of freedom of the
system $\{\zeta\}$.

It is argued that it is the volume of this system, rather
than the energy, which is the key macroscopic quantity governing
the behaviour of granular matter \cite{edwards3,edwards2}. 
If we have $N$ grains of specified shape which are
assumed to be infinitely rigid, the system's statistics would be
defined by a volume function ${\cal W}(\zeta)$, a function which gives
the volume of the system in terms of the specification of the
grains.

In this analogy  one replaces the Hamiltonian ${\cal H}(p,q)$ of
the system by the volume function, ${\cal W}(\zeta)$. The average
of ${\cal W}(\zeta)$ over all the jammed configurations determines
the volume $V$ of the system in the same way as the average of the
Hamiltonian determines the average energy $E$ of the system.

\subsection{Definition of the volume function, $\cal W$}
\label{W}

 One of the key questions in this analogy is to
establish the `correct' $\cal W$ function, the statistics of which is
capable of fully describing the system as a whole.
The idea is to partition the volume of the system  into different
subsystems, $\alpha$ with volume ${\cal W}^\alpha$, such that the
total volume of a particular configuration is
\begin{equation}
{\cal W}(\zeta) = \sum_\alpha {\cal W}^\alpha
\end{equation}

It could be that considering the volume of the first coordination
shell of particles around each grain is sufficient; thus, we may
identify the partition $\alpha$ with each grain. However,
particles further away may also play a role in the collective
system response due to enduring contacts, in which case $\cal W$
should encompass further coordination shells. In reality, of
course, the collective nature of the system induces contributions
from grains which are indeed further away from the grain in
question, but the consideration of only its nearest neighbours is
a good starting point for solving the system, and is the way in
which we proceed to describe the $\cal W$ function. The
significance of the appropriate definition of $\cal W$ is best
understood by the consideration of a response to an external
perturbation to the system in terms of analogies with the
Boltzmann equation which we will describe in Section
\ref{granularboltzmann}.


Ball and Blumenfeld \cite{ball}
have shown by a triangulation method that the area of the two-dimensional
problem can be given in terms of the contact points using
vectors constructed from them.
Here we consider a  cruder version for the volume per grain, yet with a strong
physical meaning. For a pair of grains
in contact (assumed to be point contacts for rough,
rigid grains) the grains are labelled $\alpha, \beta$, 
and 
the vector from the
centre of $\alpha$ to that of $\beta$ is denoted as 
$\vec{R}^{\,\alpha\beta}$ and 
 specifies
the complete geometrical information of the packing.  
The first step is to construct a
configurational tensor $\vec{\cal C}^{\alpha}$
associated with each grain $\alpha$
based on the structural information,

\begin{equation}
{\cal C}_{ij}^{\alpha}=\sum_{\beta} R_i^{\alpha \beta}R_j^{\alpha
\beta}.
\label{C}
\end{equation}

Then an approximation for the area in 2D or volume in 3D
encompassing the first coordination shell of the grain in question
is given as
\begin{equation} \label{Wfunc}
{\cal W}^\alpha  = 2\sqrt{\mbox{Det} {\cal C}_{ij}^\alpha}.
\end{equation}


\begin{figure}[tbp]
         \begin{center}
          \epsfig{file=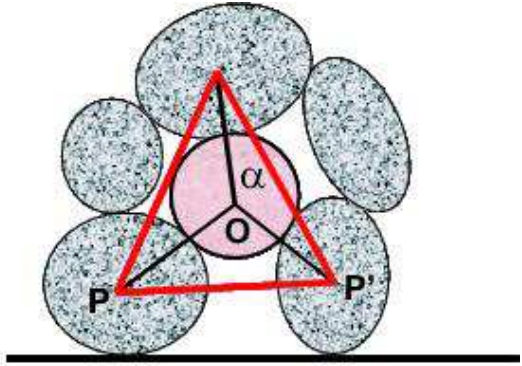,width=7 cm}
         \end{center}
\caption{Volume function $\cal W$ as discussed in the text.
\label{VOLUMES} }
\end{figure}

The volume function is depicted in the Fig. \ref{VOLUMES}, with
grain coordination number 3 in two dimensions, where Eq.
(\ref{Wfunc}) should give the area of the triangle (thick gray lines)
constructed by the centres $P$ of grains which are in contact
with the $\alpha$ grain. The
above equation is exact if the area is considered as the
determinant of the vector cross product matrix of the two sides of
the triangle. However, this definition is clearly only an
approximation of the space available to each grain since there is
an overlap of ${\cal W}^{\alpha}$ for grains belonging to the same
coordination shell. Thus, it overestimates the total volume of the
system: $\sum {\cal W}^\alpha > V$.
However, it is the simplest approximation for the system based on
a single coordination shell of a grain.

\subsection{Entropy and compactivity}
\label{entropycompactivity}

Now that we have explicitly defined $\cal W$ it is possible to
define the entropy of the granular packing.
The number of microstates for a given volume $V$ is 
measured by the area of the surface ${\cal W}(\zeta)=V$ in the
phase space of jammed configurations and it is given by:
\begin{equation}
\Sigma_{\mbox{\scriptsize jammed}}(V) = \int \delta\Big(V- {\cal
W}(\zeta) \Big)~~ \Theta(\zeta) ~~ d\zeta,
\end{equation}
where now $d\zeta$ refers to an integral over all possible jammed
configurations and $\delta(V-{\cal W}(\zeta))$ formally imposes
the constraint to the states in the sub-space ${\cal
  W}(\zeta)=V$. $\Theta(\zeta)$
is a constraint that restricts the summation to only reversible 
jammed configurations \cite{mbe}. The radical
step is the assumption of equally probable microstates
which leads to an
analogous thermodynamic entropy associated with this statistical
quantity:
\begin{equation}
\label{entropy} S(V)=\lambda \log \Sigma_{\mbox{\scriptsize
jammed}}(V) = \lambda \log \int \delta(V-{\cal W}(\zeta)) ~~
\Theta(\zeta) ~~ d\zeta,
\end{equation}
which governs the macroscopic behaviour of the system 
\cite{edwards3,edwards2}. Here
$\lambda$ plays the role of the Boltzmann constant.
The corresponding analogue of temperature,
named the ``compactivity'', is defined as

\begin{equation}\label{compactivity}
X^{-1} = \frac{{\partial S} }{{\partial V}}.
\end{equation}

\begin{figure}[tbp]
         \begin{center}
          \epsfig{file=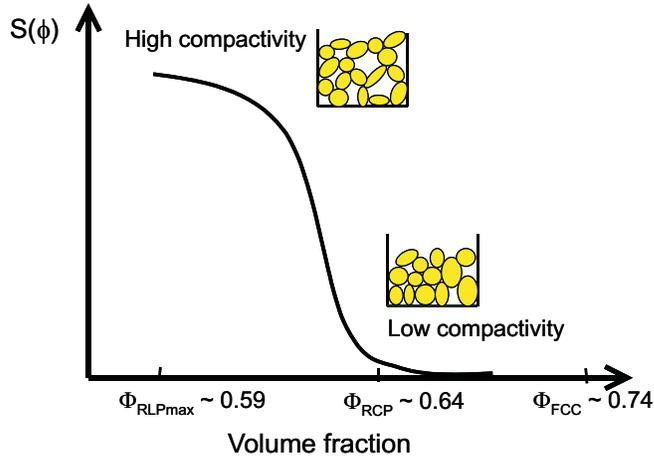,width=9 cm}
         \end{center}
\caption{ Interpretation of the compactivity and entropy in terms
of different packings.}
\label{SvsX}
\end{figure}

This is a bold statement, which perhaps requires further
explanation in terms of the actual role of compactivity in
describing granular systems. We can think of the compactivity as a
measure of how much more compact the system could be, i.e. a large
compactivity implies a loose configuration (e.g. random loose
packing, RLP) while a reduced compactivity implies a more compact
structure (e.g. random close packing, RCP). In terms of the
reversible branch of the compaction curve obtained in the Chicago 
experiments \cite{chicago2}
large amplitudes
generate packings of high compactivities, while in the limit of
the amplitude going to zero a low compactivity is achieved. In
terms of the entropy, many more configurations are available at
high compactivity, thus the dependence of the entropy on the
volume fraction can be qualitatively described as in Fig.
\ref{SvsX}. In the figure, for monodisperse packings the RCP  is
identified at $\phi\approx 0.64$, the maximum RLP
fraction is identified at $\phi\approx 0.59$, while
the crystalline packing, FCC, is at  $\phi = 0.74$ but cannot be
reached by tapping.

\subsection{Remarks}
To summarise, the granular thermodynamics is based on two
postulates:

 1) While in the Gibbs construction one assumes that
the physical quantities are obtained as an average over all
possible configurations at a given energy, the granular ensemble
consists of only the jammed configurations at the appropriate
volume.

2) As in the microcanonical equilibrium ensemble, the strong
ergodic hypothesis is that all jammed configurations of a given
volume can be taken to have equal statistical probabilities.

The ergodic hypothesis for granular matter was treated with
skepticism,
mainly because a real powder bears knowledge of its formation and
the experiments are therefore history dependent. Thus, any problem
in soil mechanics or even a controlled pouring of a sand pile does
not satisfy the condition of all jammed states being accessible to
one another as ergodicity has not been achieved, and the
thermodynamic picture is therefore not valid. 
However the Chicago experiments of tapping columns \cite{chicago2}
showed the existence of reversible situations. For instance, 
let the volume of  the column be $V(n,\Gamma)$
where $n$ is the number of taps and $\Gamma$ is the strength
of the tap. If one first obtain a volume $V(n_1,\Gamma_1)$,
and then repeat the experiment at a different
tap intensity and obtain $V(n_2,\Gamma_2)$, when we return
to tapping at $(n_1,\Gamma_1)$ one obtains a volume $V'$ which is
$V'(n_1,\Gamma_1) = V(n_1,\Gamma_1)$.
There have been several further experiments 
confirming these
results for different system geometries, particle elasticities and
compaction techniques, e. g. the system can be mechanically tapped
or oscillated, vibrated using a loudspeaker, 
or even allowed to relax under large pressures
over long periods of time, all to the same effect
\cite{bideau,sdr,cavendish}.
Moreover, in simulations of
slowly sheared granular systems 
the ergodic hypothesis was shown to work \cite{mk}.


It is often noted in the literature that although the simple
concept of summing over all jammed states which occupy a volume
$V$ works, there is no first principle derivation of the
probability distribution of the granular ensemble as it is provided
by Liouville's theorem for equilibrium statistical mechanics of
liquids and gases. In granular thermodynamics there is no
justification for the use of the $\cal W$ function to describe the
system as Liouville's theorem justifies the use of the energy in
the microcanonical ensemble. In Section \ref{granularboltzmann} we
will provide an intuitive proof for the use of $\cal W$  in
granular thermodynamics by the analogous proof of the Boltzmann
equation.

The comment was nevertheless made that there is no proof
that the entropy  Eq. (\ref{entropy}) is a 
rigorous basis for granular statistical mechanics.
Here we develop  a 
Boltzmann equation for jammed
systems and show that this analysis can be used to produce a second
law of thermodynamics, $\delta S\ge 0$ for granular matter,
and  the equality only comes with Eq. (\ref{entropy}) being
achieved.
Although everyone believes that
the second law of thermodynamics is universally true in thermal
systems, the only accessible proof comes in the Boltzmann
equation, as the ergodic theory is a difficult branch of
mathematics which will not be covered in the present discussion.
By investigating the assumptions and key points which led to the
derivation of the Boltzmann equation in thermal systems, it is
possible to draw analogies for an equivalent derivation in jammed
systems.

It should be noted that there is an extensive literature on
granular gases \cite{savage,jenkins}, which are observed when
particles are fluidised by vigorous shaking, thus inducing
continuous particle collisions. There is a powerful literature on
this topic, but it is not applicable to the problem of jamming.



\section{The Classical Boltzmann Equation}

The notion of entropy is important for thermal systems because it
satisfies the second law,

\begin{equation}
\label{secondlaw}\frac{\partial S}{\partial t} \ge 0,
\end{equation}
which states that there is a maximum entropy state which,
according to the evolution in Eq. (\ref{secondlaw}), any system
evolves toward, and reaches at equilibrium. A semi-rigorous
proof of the Second Law was provided by Boltzmann (the well-known
`H-theorem'), by making use of the `classical Boltzmann equation',
as it is now known.

\begin{figure}[tbp]
         \begin{center}
          \epsfig{file=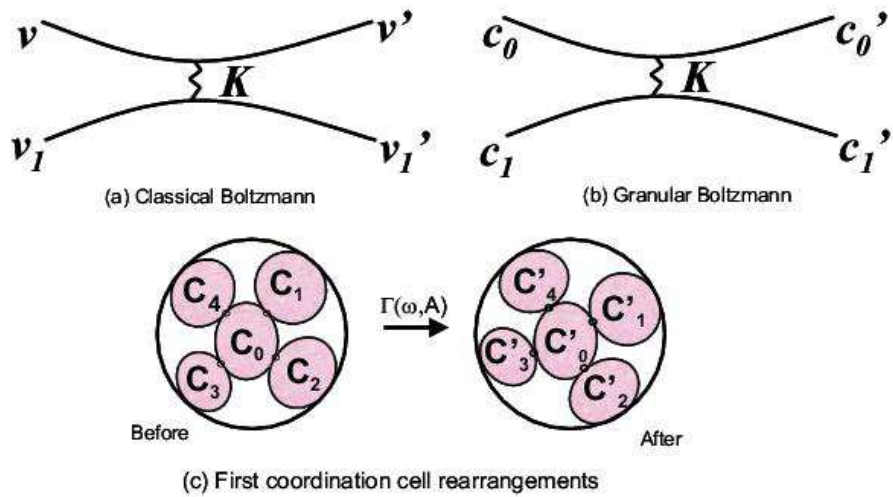,width=12 cm}
         \end{center}
\caption{(a) Collision of two particles in a dilute gas.
 (b) ``Collision of two configurations'' given in terms of
two contact points in a jammed material. (c) Rearrangements inside
a pocket ${\it a}$ under the first coordination shell
approximation of grain $\alpha=0$.\label{collision}}
\end{figure}

In order to derive this equation, Boltzmann made a number of
assumptions concerning the interactions of particles. 
The most important of these
assumptions were:

\begin{itemize}
    \item The collision processes are dominated by two-body
    collisions (Fig. \ref{collision}a).
    This is a plausible assumption for a dilute gas
since the system
    is of very low density, and the
    probability of there being three or more particles colliding
    is infinitesimal.
    \item Collision processes are uncorrelated, i.e. all memory of
the collision is lost on completion and is not remembered in
subsequent collisions: the famous Stosszahlansatz. This is also
valid only for dilute gases, but the proof is more subtle.
\end{itemize}

Thus, Boltzmann proves Eq. (\ref{secondlaw}) for a dilute gas
only, but this is a readily available situation. The remaining
assumptions have to do with the kinematics of particle collisions,
i.e. conservation of kinetic energy, conservation of momentum, and
certain symmetry of the particle scattering cross-sections.

Let $f(v,r)$ denote the probability of a particle having a
velocity $v$ at position $r$. This probability changes in time by
virtue of the collisions. The two particle collision is visualised
in Fig. \ref{collision}a where $v$ and $v_1$ are the velocities of
the particles before the collision and $v'$ and $v'_1$ after the
collision.

On time scales larger than the collision time, momentum and
kinetic energy conservation apply:

\begin{equation}
m v + m v_1 = m v' + m v_1', \\\\\\\\\\\\\\\\\\\
\frac{1}{2} m v^2 +  \frac{1}{2} m v_1^2 = \frac{1}{2} m v'^2 +
\frac{1}{2} m {v'}_1^2. \label{conservation}
\end{equation}

Then, the distribution $f(v,r)$ evolves with time according to

\begin{equation}\label{fvr}
\frac{\partial f}{\partial t}
   + v \frac{\partial f}{\partial r} +
   \int {\cal K}(v, v' ; v_1, v_1')~~ \Big (
f(v) f(v_1) - f(v') f(v_1') \Big ) ~~ d^3v_1 d^3 v'd^3v_1' = 0.
\end{equation}

The kernel $\cal K$ is positive definite and
contains $\delta$-functions to satisfy the conditions
(\ref{conservation}), the flux of particles into the collision and
the differential scattering cross-section. We consider the case of
homogeneous systems, i.e. $f=f(v)$, and define
\begin{equation}
S = - k_B \int f \log f.
\end{equation}
Defining $x = f f_1 / f' f'_1$ we obtain
\begin{equation}
\frac{\partial S}{\partial t} = \int {\cal  K}~  \log x ~(1-x)
d^3 v_1 ~ d^3 v'~ d^3v_1',\\
(1-x) \log x \ge 0, \\
{\cal K} \ge 0.
\end{equation}
Hence $ \partial S/ \partial t\ge0$ (see standard text books on
statistical mechanics).

It is also straightforward to establish the equilibrium
distribution where  $ \partial S/ \partial t = 0$ since it occurs
when the kernel term vanishes. This occurs when the condition of
detailed balance is achieved, $x=1$:
\begin{equation}\label{balance}
f(v) f(v_1) = f(v') f(v_1 ').
\end{equation}
The solution of Eq. (\ref{balance}) subjected to the condition of
kinetic energy conservation
is given by the Boltzmann distribution
\begin{equation}
\label{boltzmann} {\displaystyle f(v) = \left (\frac{k_B T}{m\pi}
\right )^{3/2} ~   e^{-\frac{1}{2} \beta m v^2} },
\end{equation}
where $\beta = 1/k_B T$.
%
Equation (\ref{boltzmann}) is a reduced distribution and valid
only for a dilute gas. The Gibbs distribution represents the full
distribution and is obtained by replacing the kinetic energy in
(\ref{boltzmann}) by the total energy of the state to obtain:

\begin{equation}
\label{gibbs} P(E) \sim  e^{-\beta E}.
\end{equation}

The question is whether a similar form can be obtained in a
granular system in which we expect

\begin{equation}
\label{canonical}
 P({\cal W}) \sim  e^{-{\cal W}/ \lambda X},
\end{equation}
where $X$ is the compactivity in analogy with
$T=\partial{E}/\partial{S}$. Such an analysis is shown in the next
section in an approximate manner.

\section{`Boltzmann Approach' to Granular Matter}
\label{granularboltzmann}

The analogous approach to granular materials consists in the
following: the creation of an ergodic grain pile suitable for a
statistical mechanics approach via a tapping 
method for the exploration of
the available configurations analogous to Brownian motion, the
definition of the discrete elements tiling the granular system via
the volume function $\cal W$, and an equivalent argument for the
energy conservation expressed in terms of the system volume
necessary for the construction of the Boltzmann equation.

We have already established the necessity of preparing a granular
system adequate for real statistical mechanics so as to emulate
ergodic conditions. The grain motion must be well-controlled, as
the configurations available to the system will be dependent upon
the amount of energy/power put into the system. This pretreatment
is analogous to the averaging which takes place inherently in a
thermal system and is governed by temperature.

As explained, the granular system explores the configurational
landscape by the external tapping introduced by the
experimentalist. The tapping is characterised by a frequency and
an amplitude ($\omega, \Gamma$) which cause changes in the contact
network, according to the strength of the tap. The magnitude of
the forces between particles in mechanical equilibrium and their
confinement determine whether each particle will move or not. The
criterion of whether a particular grain in the pile will move in
response to the perturbation will be the Mohr-Coulomb condition of
a threshold force, above which sliding of contacts can occur and
below which there can be no changes. The determination of this
threshold involves many parameters, but it suffices to say that a
rearrangement will occur between those grains in the pile whose
configuration and neighbours produce a force which is overcome by
the external disturbance.

\begin{figure}[tbp]
        \centering
       \epsfig{file=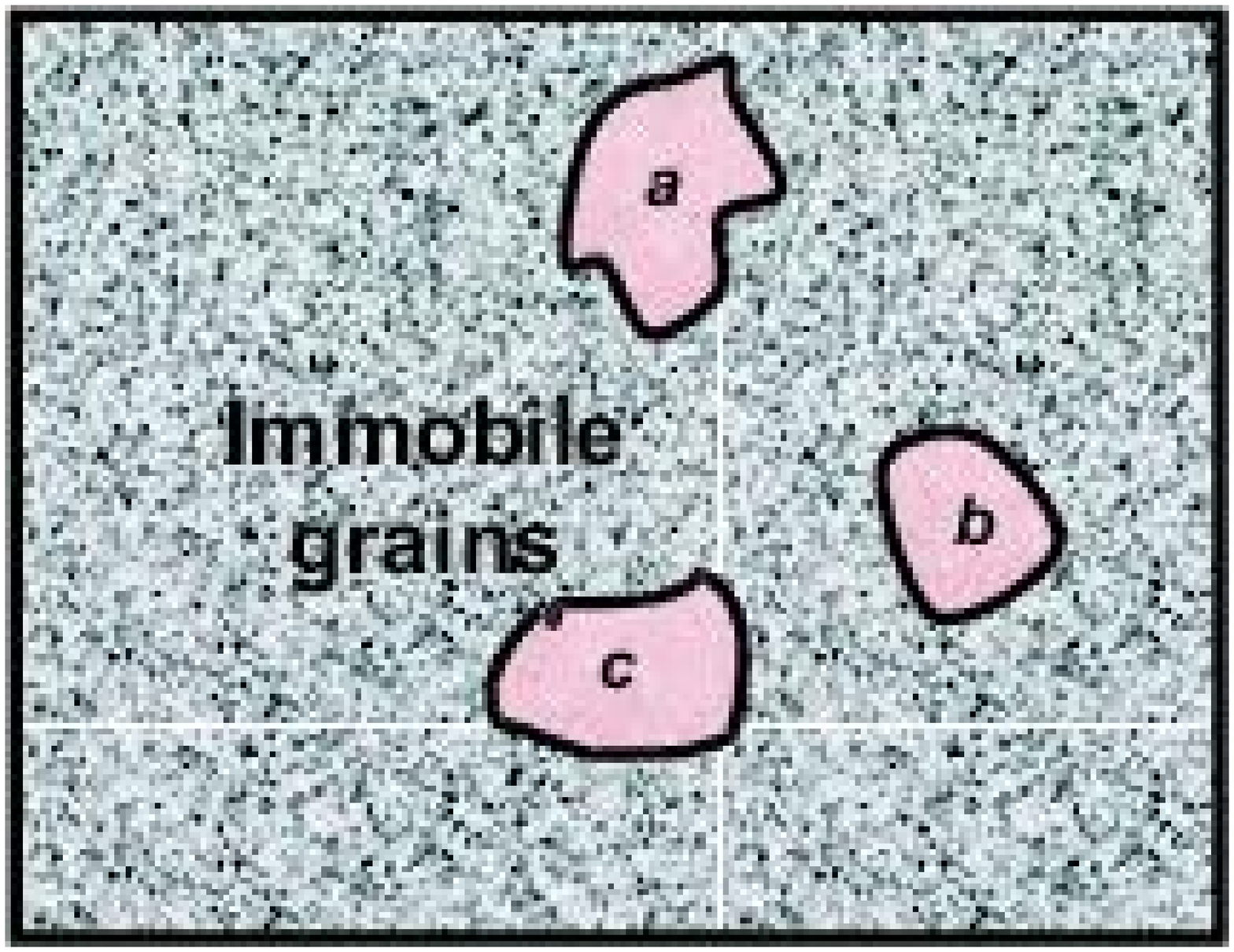,width=6 cm}
        \epsfig{file=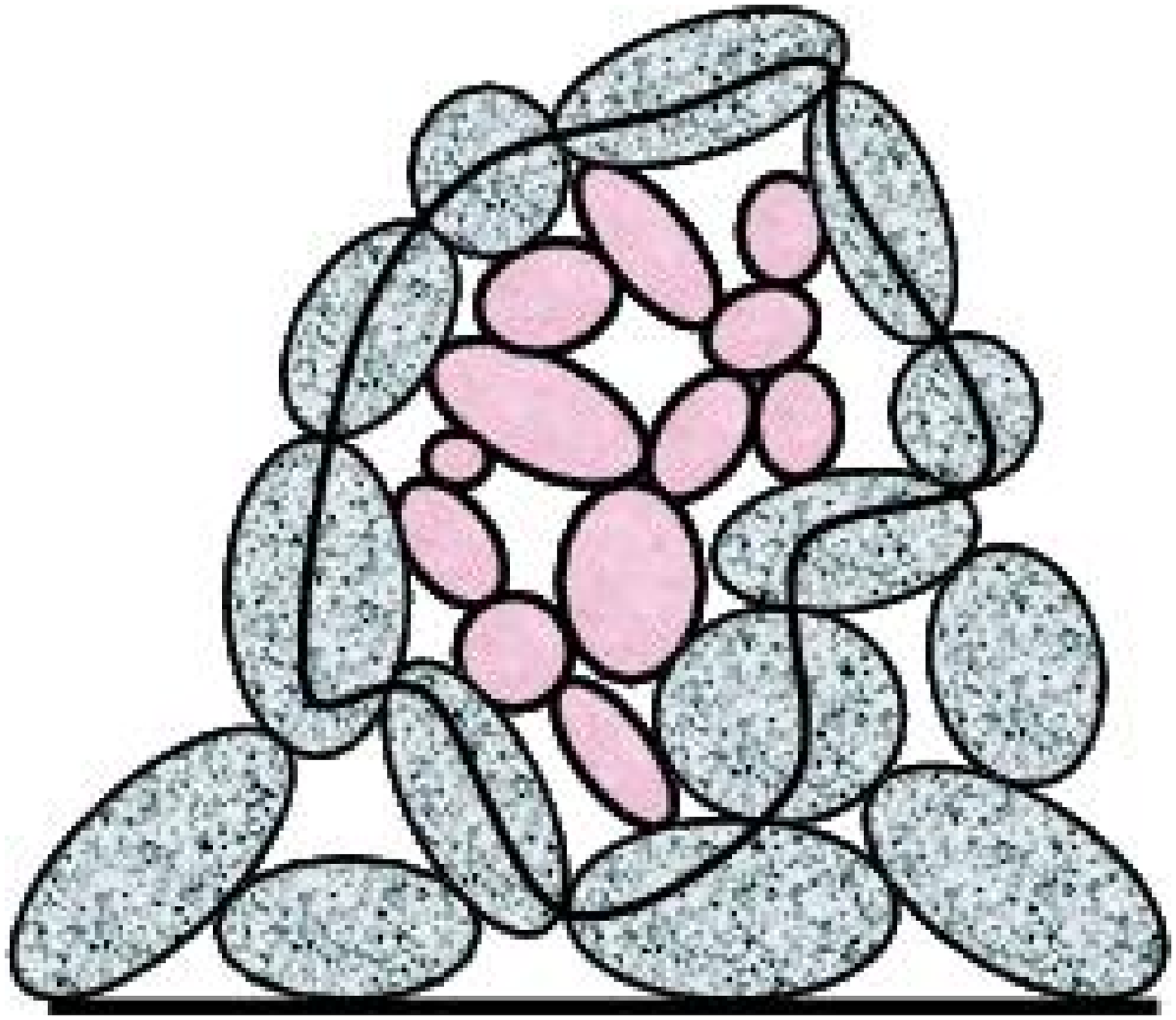,width=5 cm}
\caption{(a) Regions of mobile grains $a, b, c$ in a matrix of
immobile grains below the Coulomb threshold. (b) Detail of pocket
of mobile grains $a$ surrounded by immobile grains which are
shaded.} \label{pockets}
\end{figure}

The concept of a threshold force necessary to move the particles
implies that there are regions in the sample in which the contact
network changes and those which are unperturbed, shown in Fig.
\ref{pockets}.
Of course, since this is a description
of a collective motion behaviour, the region which can move may
expand or contract, but the picture at any moment in time will
contain pockets of motion encircled by a static matrix. Each of
these pockets has a perimeter, defined by the immobile grains. It
is then possible to consider the configuration before and after
the disturbance inside this well-defined geometry.

The present derivation assumes the existence of these regions. It
is equivalent to the assumption of a dilute gas in the classical
Boltzmann equation, although the latter is readily achieved
experimentally.


The energy input must be on the level of noise, such that the
grains largely remain in contact with one another, but are able to
explore the energy landscape over a long period of time. In the
case of external vibrations, the appropriate frequency and
amplitude can be determined experimentally for different grain
types, by investigating the motion of the individual grains or by
monitoring the changes in the overall volume fraction over time.
It is important that the amplitude does not exceed the
gravitational force, or else the grains are free to fly up in the
air, re-introducing the problem of initial creation just as they
would if they were simply poured into another container.

Within a region $a$ we have a volume $\sum_{\alpha \in a}
{\cal W}^\alpha$ and after the disturbance a volume which is now
$\sum_{\alpha \in a} {\cal W}'^\alpha$ as seen in Fig.
\ref{collision}c. In Section \ref{W} we have discussed how to
define the volume function ${\cal W}^\alpha$ as a function of the
contact network. Here the simplest ``one grain''
approximation is used as the ``Hamiltonian'' of the volume as
defined by Eq. (\ref{Wfunc}).
In reality it is much more complicated, and although there is only
one label $\alpha$ on the contribution of grain $\alpha$ to the
volume, the characteristics of its neighbours may also appear.
Instead of energy being conserved, it is the total volume which is
conserved while the internal rearrangements take place within the
pockets described above. Hence

\begin{equation}
\sum_{\alpha \in a}{{\cal W}^{\alpha}}=\sum_{\alpha \in a}{{\cal
W}'^{\alpha}} \label{volumeconserved}
\end{equation}

We now construct a Boltzmann equation. Suppose $z$ particles are
in contact with grain $\alpha=0$, as seen in Fig.
\ref{collision}c. For rough particles $z=4$ while for smooth $z=6$
at the isostatic limit. The probability distribution will be of
the contact points which are represented by the tensor ${ \cal
C}^\alpha$, Eq. (\ref{C}), for each grain, where $\alpha$ ranges from 
0 to 4 in this case.
So the analogy of $f(v)$ for the Boltzmann gas equation becomes
$f({\cal C}^0)$ for the granular system and represents the
probability that the external disturbance causes a particular
motion of the grain. We therefore wish to derive an equation

\begin{equation}
\frac{\partial f({\cal C}^0)}{\partial t}
 +    \int {\cal K}({\cal C}^\alpha,{\cal C}'^\alpha)~~ \Big (
f_0 f_1 f_2 f_3 f_4 -  f'_0 f'_1 f'_2 f'_3 f'_4 \Big )
~~d{\cal C}'^0\prod_{\alpha\neq 0} d{\cal C}^\alpha  d{\cal C}'^\alpha
= 0
\end{equation}

The term $\cal K$ contains the condition that the volume is
conserved (\ref{volumeconserved}), i.e. it must contain
$\delta(\sum {\cal W}^\alpha - \sum {\cal W}'^\alpha)$.
The cross-section is now the compatibility of the changes in the
contacts, i.e. ${\cal C}^\alpha$ must be replaced in a
rearrangement by ${\cal C}'^\alpha$ (unless these grains part and
make new contacts in which case a more complex analysis is called
for). We therefore argue that the simplest $\cal K$ will depend on
the external disturbance $\Gamma, \omega$ and on ${\cal C}^\alpha$ and
${\cal C}'^\alpha$, i.e.


\begin{equation}
\label{granularB}
\frac{\partial f({\cal C}^0)}{\partial t} +     \int
\delta \Big(\sum_\alpha {\cal
  W}^\alpha - \sum_\alpha {\cal W}'^\alpha\Big) ~ 
{\cal J}({\cal C}^\alpha,{\cal C}'^\alpha)~~ \Big (
 \prod_{\alpha=0}^{z} f_\alpha -  \prod_{\alpha=0}^{z} f'_\alpha
\Big )d{\cal C}'^0\prod_{\alpha\neq 0} d{\cal C}^\alpha  d{\cal C}'^\alpha
= 0
\end{equation}
where $\cal J$ is the cross-section and it is positive definite.

The Boltzmann argument now follows. As before
\begin{equation}
S = -\lambda \int f \log f \\ x = \frac{f_0 f_1 f_2 f_3 f_4} {f_0'
f_1' f_2' f_3' f_4'},
\end{equation}
and
\begin{equation}
\frac{\partial S}{\partial t} \ge 0,
\end{equation}
the equality sign being achieved when $x=1$ and
\begin{equation}
f_\alpha = \frac{e^{-{\cal W}^\alpha/\lambda X} }{Z}, \label{granu}
\end{equation}
with the partition function
\begin{equation}
Z = \sum e^{-{\cal W}^\alpha/\lambda X} ~~~{\Theta},
\end{equation}
and the analogue to the free energy being $Y = - X \ln Z$, and
$X=\partial V / \partial S$.

The detailed description of the kernel $\cal K $ has not been
derived as yet due to its complexity. Just as Boltzmann's proof
does not depend on the differential scattering cross section, only
on the conservation of energy, in the granular problem we consider
the steady state excitation externally which conserves volume,
leading to the granular distribution function, Eq. (\ref{granu}).

It is interesting to note that there is a vast and successful
literature of equilibrium statistical mechanics based on
$\exp(-{\cal H}/k_BT)$, but a meagre literature on dynamics based on
attempts to generalise the Boltzmann equation or, indeed, even to
solve the Boltzmann equation in situations remote from equilibrium
where it is still completely valid. It means that any advancement
in understanding how it applies to analogous situations is a step
forward.


\subsection{Experimental Validation of the Statistical Mechanics Concepts}

The first step in realising the idea of a general statistical
jamming theory
is to understand in detail the characteristics of a jammed
configuration in particulate systems. Next we present a novel
experimental method to explore this problem using confocal
microscopy \cite{behm}.

The key feature of this optical microscopy technique is that only
light from the focal plane is detected. Thus 3D images of
translucent samples can be acquired by moving the sample through
the focal plane of the objective and acquiring a sequence of 2D
images. Our model system consists of a dense packing of emulsion
oil droplets, with a sufficiently elastic surfactant stabilising
layer to mimic solid particle behaviour, suspended in a continuous
phase fluid. The refractive index matching of the two phases,
necessary for 3D imaging, is not a trivial task since it involves
unfavourable additions to the water phase, disturbing surfactant
activity. The successful emulsion system, stable to coalescence
and Ostwald ripening, consisted of Silicone oil in a solution of
water ($w_t=50\%$) and glycerol ($w_t=50\%$), stabilised by
0.01$mM$ sodium dodecylsulphate (SDS). The droplet phase is
fluorescently dyed using Nile Red, prior to emulsification. The
control of the particle size distribution, prior to imaging, is
achieved by applying very high shear rates to the sample, inducing
droplet break-up down to a radius mean size of $2\mu m$. Since the
emulsion components have different densities, the droplets cream
under gravity to form a random close packed structure. In
addition, the absence of friction ensures that the system has no
memory effects and reaches a true jammed state before measurement.

\begin{figure}[tbp]
         \begin{center}
          \epsfig{file=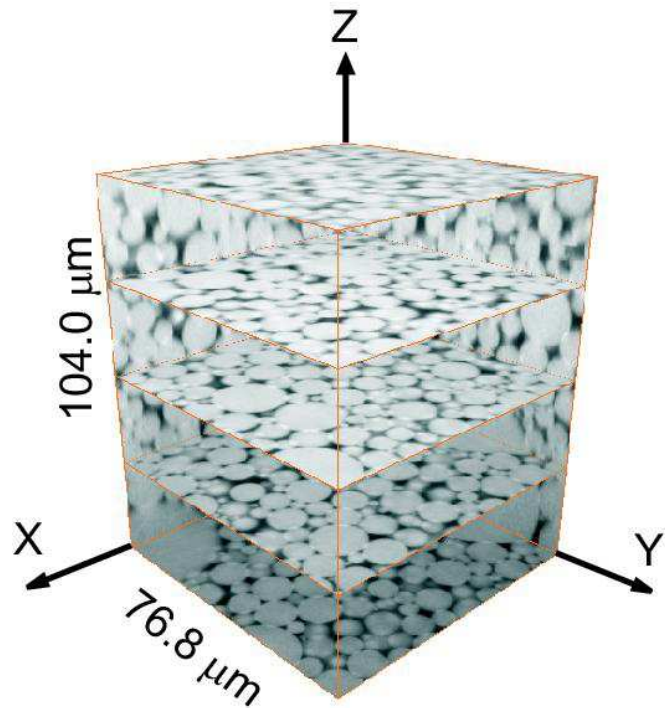,width=9 cm}
         \end{center}
\caption{Confocal image of the densely packed emulsion system.
\label{box}}
\end{figure}





\begin{figure}[tbp]
         \begin{center}
          \epsfig{file=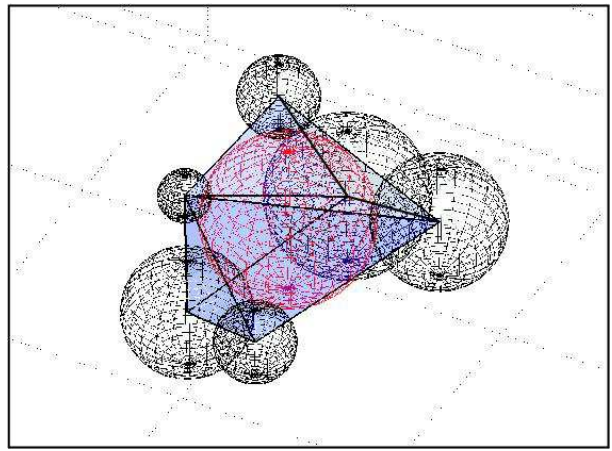,width=8 cm}
         \end{center}
\caption{An example of a volume $\cal W$ as the polyhedron
constructed from the 3D images. The centre grain (red) has 6
grains in contact (black), the centres of which are joined to form
the polyhedron. \label{Wpoly}}
\end{figure}

The 3D reconstruction of the 2D slices is shown in Fig. \ref{box}.
We have developed a sophisticated image analysis algorithm which
uses Fourier Filtering to determine the particle centres with
subvoxel accuracy. Previously, we developed a method to measure
the interdroplet forces and their distribution in the sample
volume \cite{behm}. Using an extension to the same image
analysis method, the 3D images of a densely packed particulate
model system now allow for the characterisation of the volume
function $\cal {W}$, by the partitioning of the images into first
coordination shells of each particle, described in Section
\ref{W}. The polyhedron obtained by such a partitioning is shown
in Fig. \ref{Wpoly}.

\begin{figure}[btp]
         \begin{center}
          \epsfig{file=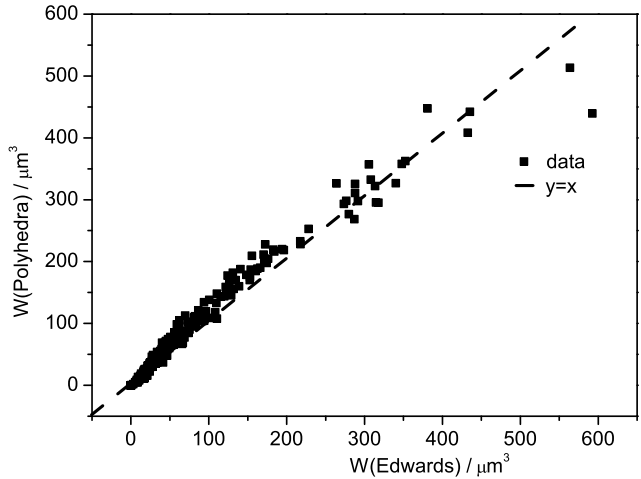,width=10 cm}
         \end{center}
\caption{$\cal W$ function obtained from the configuration tensor
$\cal C$ are plotted against the polyhedra constructed from the
images. \label{corrW}}
\end{figure}

We are able to test how the approximation in Eq. (\ref{Wfunc})
for the volume compares with the actual volume measured from the
image for each grain. Their correlation is shown in Fig.
\ref{corrW}. This approximation works well for coordination
numbers larger than 3 in 2D and even in 3D, due to the
partitioning of the obtained volumetric objects into
triangles/pyramids, intrinsic to the method, and subsequently
summing over them to obtain the resulting volume. It is clear that
very large volumes, belonging to grains with high coordination
numbers, do stray from the theoretical value due to the complex
geometries involved. According to the experimental measurements of
$\cal{W}$ employed in this study, the total volume of the system
was found to be overestimated by only $\approx5\%$.

\begin{figure}[tbp]
         \begin{center}
          \epsfig{file=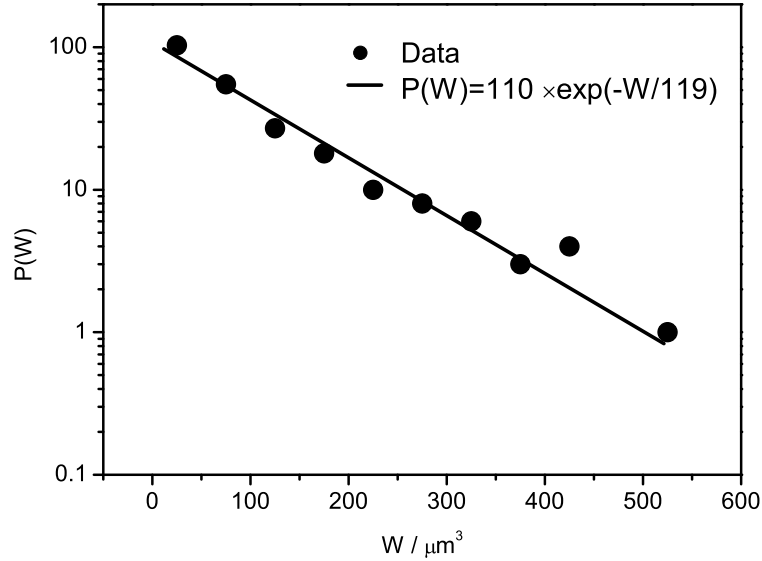,width=10 cm}
         \end{center}
\caption{Probability distribution of $\cal W$ fitted with a single
exponential. The decay constant is the compactivity, 
$X = 119\mu m^3/\lambda$. \label{distW}}
\end{figure}

The ability to measure this function and therefore its
fluctuations in a given particle ensemble, enables the
calculations of the macroscopic variables. 
In Fig. \ref{distW} we show the probability distribution of ${\cal W}$
showing an exponential behaviour as given by Eq. (\ref{granu}). 
The exponential
probability distribution of $\cal {W}$ leads to the compactivity $X$
according to Eq. (\ref{canonical}).
This implies that we can arrive at the
thermodynamic system properties from the knowledge of the
microstructure. Many images, i.e. configurations, can be treated
in this way to test whether system size influences the macroscopic
observables. If the particles are subjected to ultracentrifugation
resulting in configurations of a higher density, the influence of
pressure on the macroscopic variables can also be tested.

Such a characterisation of the governing macroscopic variables,
arising from the information of the microstructure, allows one to
predict the system's behaviour through an equation of state. This
is the first experimental study of such statistical concepts in
particulate matter and opens new possibilities for testing the
above described thermodynamic formulation. In principle, one can
apply low amplitude vibrations to the system and observe the
droplet configuration before and after the perturbation, thus
testing the ideas proposed in the Boltzmann derivation.

Acknowledgments. 
We thank D. Grinev and R. Blumenfeld for stimulating discussions.
H. Makse acknowledges financial support from the National Science 
Foundation, DMR-0239504 and the Department of Energy,
Division of Materials Sciences and Engineering, DE-FE02-03ER46089.

\end{document}